\documentstyle[11pt,paspconf,epsf]{article}

\begin{document}

\title{A Comparison of ground-based Cepheid P-L Relations with HIPPARCOS 
   Parallaxes} 
\author{H.\ Baumgardt, C.\ Dettbarn, B.\ Fuchs, J.\ Rockmann, R.\ Wielen}

\affil{Astronomisches Rechen-Institut Heidelberg, M\"onchhofstr. 12-14, 
D-69120 Heidelberg, Germany}

\begin{abstract}
The statistical test described by Wielen et al.\ (1994) is used to derive new zero-points of 
ground-based Cepheid period-luminosity (PL) and period-luminosity-colour (PLC) relations. Eleven
relations are compared with the Hipparcos data. Our results argue for a typical
increase of the adopted distance scale by about 8 \% $\pm$ 8 \%. Our zero-point for the PL relation of
Caldwell \& Laney (1991) is in agreement with that of Feast \& Catchpole (1997). 
\end{abstract}


\section{Method}

Most Hipparcos parallaxes of Cepheids are individually not very accurate and do not allow to
determine their distances with high confidence. They can nevertheless
be useful if the whole sample is used in a statistical way. For our test, photometric distances
$r_{phot}$ of all Hipparcos Cepheids are calculated according to several recent
period-luminosity and period-luminosity-colour relations. These distances $r_{phot}$ are 
converted into photometric parallaxes $\pi_{Phot}$. Figure 1 shows the differences 
between Hipparcos and photometric parallaxes $\pi_{Hipp}-\pi_{Phot}$ plotted against the 
photometric parallaxes. A linear fit of these differences according to
\begin{displaymath}
 \Delta\pi = \pi_{Hipp} - \pi_{Phot} = \pi_{0,f} + (f-1)\; \pi_{Phot}  
\end{displaymath}
provides one with a correction $\pi_{0,f}$ of the zero-points of the Hipparcos parallaxes, 
as well as with corrections $f$ of the photometric distances, where $f$ is defined as
\begin{displaymath}
 r_{phot,true} = (1/f) \; r_{phot,used} \; \; .
\end{displaymath} 

\begin{figure}
\begin{center}
\epsfxsize=10cm
\leavevmode
\epsffile{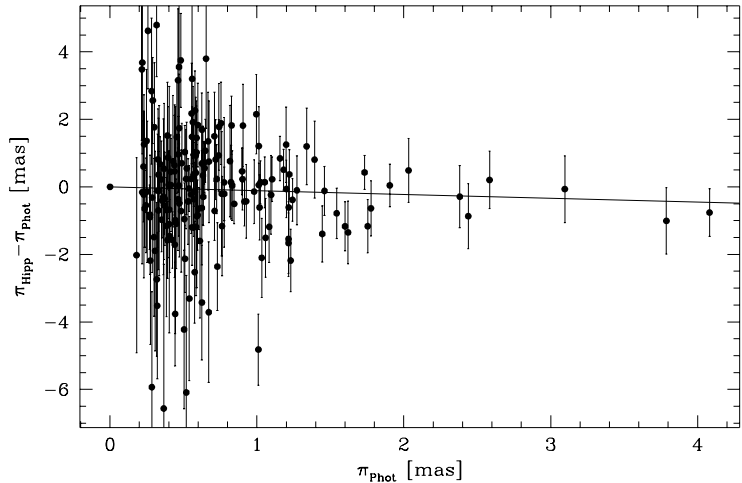}
\end{center}
\caption{Comparison of photometric parallaxes $\pi_{Phot}$ of Cepheids (based on LS94) with 
Hipparcos parallaxes $\pi_{Hipp}$. Note that the error bars represent the Hipparcos
and the photometric errors (see text).}
\end{figure}

Our method does not require to calculate individual distances from the Hipparcos parallaxes,
thereby minimizing bias of the Lutz-Kelker (1972) type. In addition,
we can utilize all stars, including those which have negative parallaxes in the Hipparcos
Catalogue. We finally note that our method avoids to take weights according to the ratio of
$\pi^2/\sigma_\pi^2$ (as in Madore \& Freedman 1998), which would introduce bias.

Caldwell \& Laney (1991) determined the spread of LMC Cepheid absolute magnitudes around
the center line of PL/PLC relations. They obtained a spread of $\sigma = 0.21$ for a PL relation
and $\sigma = 0.13$ for a PLC relation. We therefore assume uncertainties of 10\% (5\%) for the 
photometric distances derived from PL (PLC) relations. This uncertainty is transformed into an
error of the photometric parallax and we add the square of this error to the square of the error
in the Hipparcos parallax. Weights are taken according to the inverse of this sum.

\section{Sample selection}

261 stars classified as classical Cepheids were found in the Hipparcos Catalogue. From this list
all known overtone and beat Cepheids were removed. We also removed stars included in the Double and
Multiple Systems Annex of the Hipparcos Catalogue (C,G,V,O or X entries in field H59), since
the astrometric solution derived by Hipparcos may be affected by the binary nature of these 
stars.

We were left with a sample of 179 Cepheids. These are considerably fewer stars than used by Feast
\& Catchpole (1997), resulting in larger zero-point errors. This drawback is however compensated
for by the fact that our results are based on a more reliable sample of Hipparcos stars. For our 
stars Rockmann (1995) has compiled the necessary information from the literature to calculate 
their photometric distances.

\section{Results}

Table 1 lists the photometric calibrations that were compared with the Hipparcos measurements. We 
first allowed for a global zero-point error $\pi_{0,f}$ in the Hipparcos 
parallaxes. However, our test did not reveal significant zero-point errors (we typically 
obtained $\pi_{0,f} = 0.2 \pm 0.14$ for the different PL relations). 
We therefore set $\pi_{0,f}$ equal to zero. We have listed the new zero-points of 
the PL/PLC relations, obtained with $\pi_{0,f}=0$, in column 4 of Table 1.

\newcommand{\mc}[1]{\multicolumn{1}{c}{#1}}
\begin{center}
\begin{tabular}{ll@{\hspace*{1.2cm}}lcc}
\noalign{\smallskip}
\hline
\hline
\noalign{\smallskip}\\[-0.08cm]
\hspace*{-0.15cm}Author(s) & \hspace*{-0.2cm}
Type &\hspace*{-0.8cm}Old Zero-Point &New Zero-Point & 1/f\\[+0.2cm]
\noalign{\smallskip}
\hline
\noalign{\medskip}
FW87  & PLC & $-2.27$  & $-2.58\; \pm\; 0.16$ & $1.15\; \pm\; 0.08$\\ 
GBM93 & PL  & $-1.32$  & $-1.30\; \pm\; 0.17$ & $0.99\; \pm\; 0.08$\\
GF93  & PL  & $-1.30$  & $-1.43\; \pm\; 0.17$ & $1.06\; \pm\; 0.08$\\
LS94  & PL  & $-1.197$ & $-1.45\; \pm\; 0.17$ & $1.13\; \pm\; 0.08$\\
IT75  & PLC & $-2.77$  & $-2.67\; \pm\; 0.16$ & $0.95\; \pm\; 0.08$\\
CL91A & PLC & $-2.01$  & $-2.33\; \pm\; 0.16$ & $1.16\; \pm\; 0.08$\\
HB89  & PL  & $-1.542$ & $-1.26\; \pm\; 0.17$ & $0.88\; \pm\; 0.08$\\
MF91  & PL  & $-1.39$  & $-1.54\; \pm\; 0.17$ & $1.07\; \pm\; 0.08$\\
B87   & PL  & $-1.24$  & $-1.52\; \pm\; 0.17$ & $1.14\; \pm\; 0.08$\\
CL91B & PL  & $-1.31$  & $-1.50\; \pm\; 0.17$ & $1.09\; \pm\; 0.08$\\
GFG98 & PL  & $-1.294$ & $-1.54\; \pm\; 0.17$ & $1.12\; \pm\; 0.08$\\
\noalign{\smallskip}
\hline
\noalign{\smallskip} 
\end{tabular}
\end{center}
{\it Table1: Photometric calibrations compared with the Hipparcos parallaxes. The new zero-points and their errors
can be found in column 4. The abbreviations in column 1 have the following meaning: FW87: Feast \&
Walker (1987); GBM93: Gieren, Barnes \& Moffet (1993); GF93: Gieren \& Fouqu\'e (1993); 
LS94: Laney \& Stobie (1994); IT75: Iben \& Tuggle (1975); CL91A: Caldwell \& Laney (1991); 
HB89: Hindsley \& Bell (1989); MF91: Madore \& Freedman (1991); B87: Berdnikov (1987);
CL91B: Caldwell \& Laney (1991); GFG98: Gieren, Fouqu\'e \& G\'omez (1998)}\\

The Hipparcos parallaxes indicate that the zero-points of most Cepheid PL/PLC relations have to 
be shifted by 0.0 - 0.3 mag to brighter magnitudes, indicating that the previous distance scales 
were too short by about 10~\%. For most relations, the shifts are however of the same order as the 
errors.

We obtain a new zero-point of $\rho = -$1.50$\pm 0.17$ for the PL relation of Caldwell \& Laney 
(1991B). This agrees very well with the value $\rho = -1.43 \pm 0.10$, that Feast \& 
Catchpole (1997) determined as the new zero-point of this PL relation. Values between $\rho =
 -$1.52$\pm 0.13$ and $\rho = -$1.45$\pm 0.13$, depending on the adopted reddening, were 
also derived by Feast et al.\ (1998). If we compare our zero-point with the apparent zero-point of 
the LMC Cepheids from Caldwell \& Laney (1991) ($\rho = 17.23$), and add a metalicity-correction of 
+0.042 (Laney \& Stobie 1994) to the difference, we obtain a distance modulus of $18.77 \pm 0.17$ 
to the LMC.

The PL relation of Madore \& Freedman (1991) is of particular interest since the HST Key Project
on the Extragalactic Distance Scale uses this relation to determine the distances to other 
galaxies containing Cepheids. For their relation, we derive a distance scale correction of 
$1/f = 1.07 \pm 0.08$, i.e.\ an increase of all distances and a corresponding decrease of the 
Hubble Constant by 7\% $\pm$ 8 \%.

Monte-Carlo simulations were performed in order to check the influence of Malmquist bias on our
results. We assumed a uniform spatial distribution of the Cepheids and a spread of their absolute 
magnitudes 
of 0.2 mag around the center line. The distribution of the Cepheids over periods and the 
completeness function of our sample were varied. The resulting bias in our zero-point  
was in most cases below 0.01 mag and never exceeded 0.015 mag. We therefore conclude that  
Malmquist bias plays only a minor r\^ole and have neglected it in our zero-point determination. 

We finally note that our results do not change significantly if the nearby Cepheids are omitted from 
our analysis. For example, restricting our analysis to Cepheids with photometric distances above 
500 pc would change the zero-point of the PL relation of Laney \& Stobie (1994) by only 0.04 mag to
$-1.41 \pm 0.21$. Shifts of the same order are also obtained for the other PL/PLC relations. We 
conclude that our changes for the zero-points are not due to a few nearby (and perhaps peculiar)  
Cepheids.\\

\begin{acknowledgements}

We are grateful to Fr\'ed\'eric Arenou for useful discussions. HB is supported by the 
Sonderforschungsbereich 328 Entwicklung von Galaxien.
\end{acknowledgements}

\end{document}